\documentclass{aa}
\usepackage{graphicx}
\usepackage{url}

\usepackage{txfonts}
\usepackage[breaklinks,colorlinks,citecolor=blue]{hyperref}
\usepackage{natbib,twoopt}
\newcommand{\GG}[1]{}

\bibpunct{(}{)}{;}{a}{}{,} 
\makeatletter
\newcommandtwoopt{\citeads}[3][][]{\href{http://adsabs.harvard.edu/abs/#3}%
{\def\hyper@linkstart##1##2{}%
\let\hyper@linkend\@empty\citealp[#1][#2]{#3}}}
\newcommandtwoopt{\citepads}[3][][]{\href{http://adsabs.harvard.edu/abs/#3}%
{\def\hyper@linkstart##1##2{}%
\let\hyper@linkend\@empty\citep[#1][#2]{#3}}}
\newcommandtwoopt{\citetads}[3][][]{\href{http://adsabs.harvard.edu/abs/#3}%
{\def\hyper@linkstart##1##2{}%
\let\hyper@linkend\@empty\citet[#1][#2]{#3}}}
\newcommandtwoopt{\citeyearads}[3][][]%
{\href{http://adsabs.harvard.edu/abs/#3}
{\def\hyper@linkstart##1##2{}%
\let\hyper@linkend\@empty\citeyear[#1][#2]{#3}}}
\makeatother

\newcommand{\msun}{$\rm M_{\odot}$ }

\def \deg         {\text{$^{\circ}$}}

\begin{document}

   \title{Clues on the nature of the quasi-periodic optical outbursts of the blazar OJ 287}

   \titlerunning{Quasi-periodic optical outbursts of OJ 287}

\author {Gopal-Krishna\inst{1}\thanks{E-mail: gopaltani@gmail.com}
}
  
\institute{$^{1}$UM-DAE Centre for Excellence in Basic Sciences (CEBS), Vidyanagari, Mumbai - 400098, India\\
}

 \date{\today} 
 
 \abstract
{The fascination with the blazar OJ 287 stems not only from its status as 
a prominent candidate for a close supermassive black hole (SMBH) binary, but also because of the thermal bremsstrahlung origin proposed for its giant optical outbursts. These outbursts arrive as pairs, quasi-periodically every $\sim 12$ years, based on the unique 130-year-long, well-sampled optical light curve available for this blazar. For its three well-known, large quasi-periodic optical outbursts (QPOOs), observed in 1983, 2007 and 2015, optical photo-polarimetric monitoring has been reported in the literature. For these initially radio-undetected QPOOs, widely acclaimed as `bremsstrahlung flares', we have scrutinised the available measurements of optical polarisation 
and spectral index during the rising phase. Several inconsistencies of these data with the optical bremsstrahlung interpretation are noted, which point towards a synchrotron-dominated alternative interpretation for all these prominent QPOOs, just as for the optical emission observed between the outbursts. Possible reasons for the radio non-detection of the QPOOs during the initial stage are outlined.
} 


\maketitle

\section{Introduction} \label{sec:intro}

Even among the tiny sub-population of active galactic nuclei (AGNs) known as blazars, exhibiting strong variability, OJ 287 (J0854+2006, PG0851+202, $z$ = 0.306) stands out for its highly intriguing optical variability patterns. Its exceptionally long recorded optical light curve, going back to 1888, is dotted with outbursts which occur with a quasi-periodicity of about 12 years \citep[][]{Sillanpaa1988ApJ...325..628S,Hudec2013A&A...559A..20H}. These outbursts  are double-peaked with a peak separation of order of one year \citep[][]{Sillanpaa1996aA&A...315L..13S}.
Between such quasi-periodic optical outbursts (QPOOs), the observed optical emission, as in the case of all known blazars, is predominantly of synchrotron origin, as inferred from its generally high polarisation levels \citep[e.g.][]{Darcangelo2009ApJ...697..985D,Villforth2009MNRAS.397.1893V,Rakshit2017ApJ...835..275R,Singh2022aJApA...43...85S,Gupta2023ApJ...957L..11G}, strong intranight optical variability (INOV,
\citealp[e.g.][]{Sillanpaa1991AJ....101.2017S,Sagar2004MNRAS.348..176S,Villforth2009MNRAS.397.1893V,Goyal2013MNRAS.435.1300G,Rakshit2017ApJ...835..275R,Gupta2017MNRAS.465.4423G,Wehrle2019ApJ...877..151W}) and SED modelling \citep[e.g.][]{Abdo2009ApJ...700..597A,Kushwaha2018aMNRAS.479.1672K}. The latter demonstrated it to be a low-frequency-peaked blazar \citep[][]{Padovani1995ApJ...444..567P}.
It had been proposed by \citet[][]{Sillanpaa1988ApJ...325..628S} that the QPOOs originate from a close binary of supermassive black holes (SMBHs) undergoing general relativistic precession, such that the pericentric passage of the secondary SMBH triggers a tidally induced mass flow from the disk to the primary SMBH  \citep[see, also][]{Valtaoja2000ApJ...531..744V}. Another major difference from typical blazars is that the dominant starting flare of the QPOOs is not detected at radio wavelengths \citep[e.g.][]{Sillanpaa1996bMmSAI..67..539S,Valtaoja2000ApJ...531..744V}. Moreover, it was claimed to be accompanied by a decreasing percentage optical polarisation, $P$, \citep[see reviews by][]{Takalo1994VA.....38...77T,Valtonen2016ApJ...819L..37V,Kushwaha2022JApA...43...79K,Valtonen2023aGalax..11...82V}. The concomitant drop in $P$, together with the radio non-detection, led to the `bremsstrahlung flare' paradigm for the QPOOs \cite[e.g.][]{Sillanpaa1991Ap&SS.176..297S,Lehto1996ApJ...460..207L,Valtaoja2000ApJ...531..744V} and to the widely explored `disk-piercing’ (DP) model of OJ 287, introduced by \citet[][]{Lehto1996ApJ...460..207L}, specifically to explain the arrival of the QPOOs in pair, with a year-like separation. In this DP model, the orbiting secondary SMBH, upon impacting the disk of an ultra-massive primary SMBH ($\sim 10^{10}$ \msun), punches a hole in the disk, twice per its highly eccentric, fast-precessing orbit. An optical outburst (`impact flare’) of bremsstrahlung radiation is supposed to ensue at each piercing of the disk by the secondary SMBH, as the shock-heated disk plasma is pulled out of the impact site and the plasma bubbles blown supersonically on either side of the disk expand and cool \citep[see, also][]{Ivanov1998ApJ...507..131I,Pihajoki2013aMNRAS.434.3122P,Valtonen2019ApJ...882...88V}. 
Within this framework, the expected gravitational response of the accretion disk to the orbiting secondary SMBH has been calculated by \citet[][]{Sundelius1997ApJ...484..180S} and \citet[][]{Pihajoki2013aMNRAS.434.3122P} and it constitutes an essential ingredient to predicting the arrival times of the QPOOs \citep[e.g.][and references therein]{Valtonen2023aGalax..11...82V}. We recall that the DP model posits a geometrically thin (Shakura-Sunyaev) accretion disk; however, this assumption was recently questioned by \citet {Antonucci2023Galax..11..102A}\citep[see also,][]{Villforth2010MNRAS.402.2087V}.

Variants of the binary SMBH scenario for explaining the QPOOs of this blazar, some of which associate the QPOOs directly with the primary SMBH, include those proposed by 
\citet[][]{Sillanpaa1988ApJ...325..628S}, \citet{Katz1997ApJ...478..527K}, \citet{Villata1998MNRAS.293L..13V}, \citet{Valtaoja2000ApJ...531..744V},  \citet{Liu2002A&A...388L..48L}, \citet{Qian2015RAA....15..687Q}, \citet{Dey2018ApJ...866...11D}, \citet{Britzen2018MNRAS.478.3199B}, and \citet{Britzen2023ApJ...951..106B}. In particular, the latter two papers have explicitly argued for a substantial deterministic component in the observed variability of flux density and of the SED state of OJ 287 \citep[see,][]{Kushwaha2018aMNRAS.479.1672K} by proposing a model of precession and nutation jet motions, partly constrained by the available four-decade-long sequence of very long baseline interferometry (VLBI) images \citep[see also][]{Cohen2017Galax...5...12C,Butuzova2020Univ....6..191B}. 
We note that these models belong to a category where the binary is not required to have an ultra-massive black hole ($\sim 10^{10}$ \msun). To verify the predictions of the different models, a dedicated long-term program of dense monitoring of OJ 287 at over a dozen different frequencies from radio to X-rays has been carried out since late 2015: Multi-wavelength Observations and Modelling of OJ 287 (MOMO, \citealp[][]{Komossa2021bUniv....7..261K}). The project makes use of a large number of ground-based observatories worldwide and space observatories. This has made OJ 287 one of the best-monitored blazars, particularly in the X-ray/UV/optical regimes \citep[][]{Komossa2023aApJ...944..177K,Komossa2023bAN....34420126K}.

The goal of the present study is basically limited to investigating the nature of the radiation associated with a set of well-observed, prominent QPOOs of OJ 287, which are much-acclaimed cases of 'bremsstrahlung flares' \citep[e.g.][]{Valtonen2021Galax..10....1V}. In Sect. 2, we carried out consistency checks on three such QPOOs, observed in 1983, 2007 and 2015, with the hypothesis of their  bremsstrahlung origin, which predicts: (i) a flattening of the optical spectral index ($\alpha$) due to dilution by the thermal emission, whose spectrum should be nearly flat; (ii) a dip in the optical fractional polarization ($P)$, again due to dilution by the thermal emission which is expected to be unpolarised; and (iii) general stability (i.e. the absence of large fluctuations and/or swings) of the polarisation angle ($\theta$), since $\theta$ cannot be altered by addition of the (unpolarised) bremsstrahlung emission \citep[e.g. see][]{Valtaoja2000ApJ...531..744V}. Thus, these are the tell-tale signs of a bremsstrahlung flare, which can be looked for in the existing optical data. Among these, we deem the $\theta$-test to be the most powerful test of the thermal dilution hypothesis. This is because (unlike the first two tests) the $\theta$-test is practically immune with respect to the uncertainty about the level of the underlying synchrotron emission during a putative bremsstrahlung flare, a factor that weakens the other two tests. For instance, this uncertainty virtually precludes the error estimation for the derived spectral index \citep[see e.g.][]{Valtonen2023aGalax..11...82V}. Furthermore, the error gets accentuated due to the small factor of just $\sim$ 2 in frequency, which the optical window affords. 

The aforementioned three famous QPOOs chosen here for the purposes of the above tests are an unbiased selection, guided solely by the availability of published optical photo-polarimetric measurements covering several consecutive ( > 5) nights around the QPOO peak. We recall that a bremsstrahlung flare in the DP model is attributed to collision of the orbiting secondary SMBH with the disk around the (much more massive) primary SMBH. In this process, a bubble of shock-heated plasma is released out of the disk on each side and envelopes the impact site as it expands and cools adiabatically (see above).\footnote{\citet[][]{Tanaka2013MNRAS.434.2275T} has suggested an alternative model for the QPOOs, wherein the secondary SMBH creates a cavity in the disk once every orbit and the disk plasma leaks into it. In a radically different approach which does not involve a SMBH binary, \citet[][]{Villforth2010MNRAS.402.2087V} have proposed a `magnetic breathing' scenario to explain the double-peaked optical outbursts of OJ 287.} 
As the expanding plasma bubble becomes transparent, it is posited to emit a burst of `bremsstrahlung’ radiation, often capable of overwhelming the underlying synchrotron radiation. After some further cooling, the thermal plasma bubble would evolve into an H II region \citep[e.g.][]{Valtonen2021Galax..10....1V}. We also note that the QPOOs are found to be sharply peaked initially and compared to the rising phase, which lasts just a couple of days, the decay is slower, probably due to an observed surging contribution from (polarised) optical synchrotron emission, possibly originating near the impact site and becoming increasingly prominent as the opacity of the thermal plasma bubble declines due to its continued expansion \citep[e.g.][]{Valtonen2021Galax..10....1V}.
For this reason, it is much preferable to look for signatures of thermal dilution by a putative `bremsstrahlung' component during the rising part of the QPOO, where this component is supposed to dominate according to the DP model. This is the approach followed below.

\section{Analysis of the photo-polarimetric monitoring observations of the three prominent QPOOs}

The three QPOOs of OJ 287, with published optical photo-polarimetric measurements on several consecutive nights around the flux maximum, are associated with: January 1983 \citep[][]{Smith1987ApJS...64..459S}, September 2007 \citep[][]{Valtonen2008Natur.452..851V}, and December 2015 \citep[][]{Valtonen2016ApJ...819L..37V}. For the first two QPOOs, the monitoring data of polarisation position angle ($\theta$) have also been reported, in addition to the fractional polarisation $(P)$ and the flux monitoring data. We examine these three datasets below, proceeding in reverse chronological order. 

\subsection{The outburst of 5 December 2015: }

This large QPOO began around mid-November 2015, coinciding with the centenary of Einstein’s  general relativity theory; hence, its denomination as a `GR’ flare \citep[e.g.][]{Valtonen2016ApJ...819L..37V}. It was the brightest QPOO observed since the January 1983 QPOO (Sect. 2.3). The outburst consisted of two main flares, altogether spanning $\sim$ 40 days. The first (larger) flare peaked on 5 December 2015 (JD 2457361). The second (strongly polarised) flare, peaking $\sim$ 18 days later, is clearly a synchrotron flare, thought to originate from the same impact site \citep[e.g.][]{Valtonen2019ApJ...882...88V}. The published data allow for the following two tests to be carried out:

{\bf (a) The $\alpha$-test:} The available multi-colour photometry \citep[][]{Gupta2017MNRAS.465.4423G} has scanty pre-flare coverage. Nonetheless, it is seen from their Fig. 2 that the (V-R), (B-V), and (R-I) colour indices at the flare maximum are very close to their long-term values and any flattening of $\alpha$ at the flare maximum is $< $ 0.05. According to these authors, $\alpha_{VR}$ for the low brightness state was $\sim$ -2.0, while for the outburst (QPOO) state $\alpha_{VR}$ was $\sim$ -2.13. Thus, there is no evidence of $\alpha$ flattening during the outburst.\\

{\bf (b) The $P$-test:} Fig. 3 of \citet[][]{Valtonen2016ApJ...819L..37V} displays measurements of $P$ (R-band) during this QPOO. For 5 consecutive days up to the `bremsstrahlung' peak (JD 2457361), the plot shows 8 measurements of $P$, out of which four were made using 2-meter class telescopes and the remaining four with a 0.6-metre telescope (these latter four values are given below in parentheses).These measurements of $P$, read from their Fig. 3 are, starting from the day of the peak: [6.3\%, (8.8)\%], -1 day [(5.6\%), 7.4\%, 7.0\%], -2 day [(10.8\%)], -3 day [(2.8\%)], and -4 day [6.7\%].
The corresponding averaged values of $P$ are 7.6\% (on the day of the flare peak), 6.7\% (-1 day), 10.8\% (-2 day), 2.8\% (-3 day), and 6.7\% (-4 day). We note that during those 5 days, the flux of the flare rose $\sim 3$ times. If we now confine attention to the four measurements made using 2-m class telescopes, the values of $P$ are: 6.3\% (at the flare maximum), 7.2\% (-1 day), and 6.7\% (-4 day). Thus, either way, no systematic decline in $P$, in sync with the nearly three-fold steady rise in flux, is evident.
Unfortunately, the $\theta$-test for this QPOO is precluded due to 
non-availability of $\theta$ monitoring data\footnote{Actually, $\theta$ light curve for this QPOO has recently been published by that team (see, Fig. 1 of \citet[][]{Gupta2023ApJ...957L..11G}. However, individual data points in their long-term $\theta$ 
light curve cannot be visually separated in time, because of the heavy compression of the plot along the time axis. Also, the date of the `bremsstrahlung' flare maximum is given by them as 8 December 2015, which is three days after the date of 5 December 2015 given in \citet[][Sect. 2.1]{Valtonen2016ApJ...819L..37V}. The difference is more than the rise time-scale of the flare ($\sim$ 2 days). Origin of this discrepancy is unclear.}

\subsection{The outburst of 13 September 2007: }
{\bf (a) The $P$-test:} Polarimetric measurements of this QPOO are displayed in Fig. 2 of \citet[][]{Valtonen2008Natur.452..851V}. 
These cover three consecutive nights preceding the continuum peak (13 September 2007) and three consecutive nights following the peak. The observed two-fold decline in $P$ (starting from $P \sim 8\%$), concurrently with the doubling of the flux seems consistent with the thermal dilution hypothesis. However, intriguingly, the maximum dip in $P$, taking it down to $\sim 2\%$, occurred two nights before the continuum peak{\footnote{Within the synchrotron jet scenario, a possible explanation for such a temporal offset between flux and $P$ has been provided in the `swinging jet’ model 
\citep[][]{Gopal-Krishna1992A&A...259..109G}.}
\citep[see,][]{Valtonen2008Natur.452..851V,Valtonen2011AcPol..51f..76V}; {\bf (b) The $\theta$-test:} Fig. 2 of \citet[][]{Valtonen2008Natur.452..851V}, shows that the only large inter-night spike in $\theta$ (by $\sim 35\deg$) within the 40 days long plot, occurred precisely on the night of the QPOO peak. Clearly, both these polarimetric results are difficult to reconcile with the bremsstrahlung interpretation of this flare.

\begin{figure*}
\centering
\includegraphics[scale=0.52]{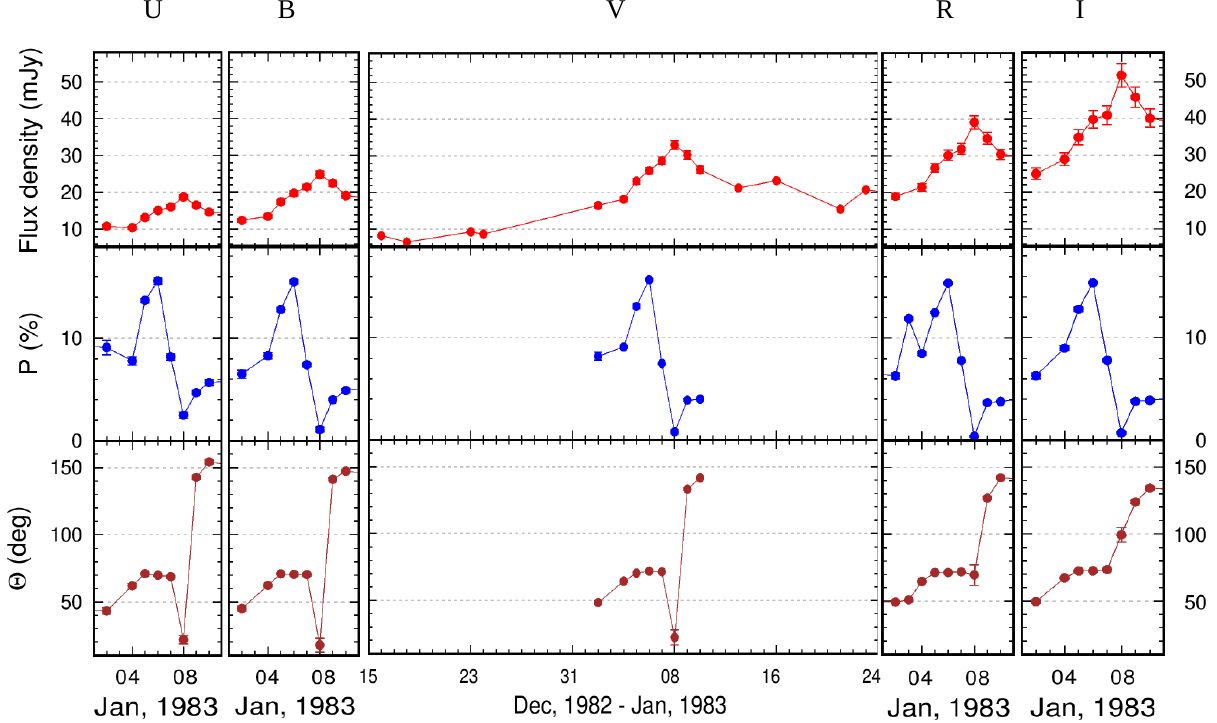} 
\caption{UBVRI photo-polarimetric measurements of the January 1983 QPOO of OJ287. The data are taken from \citet[][]{Smith1987ApJS...64..459S}. In the V-band light curve, the first and the last four data points are taken from \citet[][]{Sillanpaa1991Ap&SS.176..297S}}. 
\label{fig:LC}
\end{figure*}

\subsection{The giant outburst of 8 January 1983:}

For this QPOO, the available data \citep[][]{Smith1987ApJS...64..459S,Sillanpaa1991Ap&SS.176..297S} permit all three tests of the bremsstrahlung hypothesis:

{\bf (a) The $\alpha$-test:}
Based on the V-band light curve of the QPOO, displayed in their Fig. 1, \citet[][]{Valtonen2008Natur.452..851V} set 1983.00 as the starting point of this QPOO. The flare is seen to peak on 8 January, before fading (albeit, dotted with a few smaller synchrotron flares) back to the pre-flare level in $\sim$ 45 days. The rising part of the flare amounts to at least a three-fold increase in flux density (Fig. \ref{fig:LC}), which \citet[][]{Valtonen2008Natur.452..851V,Valtonen2021Galax..10....1V} identified as a bremsstrahlung event.
Such a massive infusion of thermal flux (whose optical spectrum is expected to be nearly flat) should have resulted in a conspicuous spectral flattening in tandem with the flux rise during this giant flare. However, the observed spectral flattening during this week-long rising phase was very marginal ($\Delta\alpha < 0.05$, see Fig. 2(b) of \citet[][]{Sillanpaa1991Ap&SS.176..297S}. This is clearly difficult to reconcile with the thermal dilution, namely, the bremsstrahlung hypothesis. 

{\bf (b) The $P$-test: } The observed drop in $P$ near the peak of this giant QPOO has long been  cited as powerful evidence of a bremsstrahlung-dominated rise of the QPOOs of OJ 287 \citep[e.g.][ however, see more details below.]{Valtonen2021Galax..10....1V}
From Fig. \ref{fig:LC}, we can see that the light curves in all five optical bands exhibit a general rise in flux density over the five days preceding the peak on 8 January, whereafter a similar decline is seen over the next three days, becoming more gradual over the following week. At the first sight, the observed dipping of $P$ to a minimum exactly on the night of the continuum peak (8 January) does appear to be in accord with the hypothesis of thermal dilution of the existing synchrotron flux, by the flare. However, we argue below that in quantitative terms, this inference is too simplistic.

From Fig. \ref{fig:LC}, the sharpest drop in $P$ occurred between January 6-8 and we now examine if this precipitous decline in $P$ was indeed due to thermal dilution. From Table 3 of \citet[][]{Smith1987ApJS...64..459S}, we see that the R-band flux densities on these two dates were: 30.2$\pm$1.4 mJy and 39.2$\pm$1.8 mJy (the peak). The corresponding observed values of $P$ were 15.4$\pm$0.1\% and 0.4$\pm$0.1\%. This plunge in $P$, which must be real since it is seen in all the 5 bands (Fig. \ref{fig:LC}), took place as $\sim$ 9 mJy were concurrently added to the flare (R band). Even if all this added flux were bremsstrahlung emission, and making a neutral assumption that the underlying synchrotron component experienced no large change over the two-day interval, the observed increase in flux from $\sim 30$ to $\sim$ 39 mJy, would have only decreased $P$ from the observed $15.4\pm 0.1\%$ on 6 January to $\sim 12\%$ on 8 January, which is 30 times higher than $P = 0.4\pm0.1\%$ actually measured on 8 January. Clearly, most of the observed fall in $P$ did not occur due to thermal dilution and probably has a different origin, possibly a rapid variation in the flux/polarisation of the underlying synchrotron component \citep[e.g. see][]{Holmes1984MNRAS.211..497H}. This would not be so surprising for blazars are known to exhibit large (and often correlated) rapid changes in optical continuum and polarisation, thanks to both arising from the same relativistically beamed synchrotron jet. Thus, once again, this QPOO seems to be a synchrotron-dominated event and any role played by thermal dilution is secondary. \\ 

{\bf (c) The $\theta$-test:} As mentioned in Sect. 1, this is probably the cleanest test of the thermal dilution hypothesis. Thus, although thermal dilution is expected to leave $\theta$ unaffected, we notice from Fig. \ref{fig:LC} here, along with the V-band data given in Table 3 of \citet[][]{Smith1987ApJS...64..459S}, that a large drop in $\theta$ occurred on the night of the flare maximum (8 January): from 71.4$\pm$0.5 deg on 7 January to 22.3$\pm$5.6 deg on 8 January ( 8.7$\sigma$ change) and then to 133.2$\pm$1.0 deg on the very next night (20$\sigma$ change). Likewise, between 7-9 January, $\theta$ changed by 61.8$\pm$1.1 deg (55$\sigma$). All this leaves little doubt that the giant QPOO of January 1983 was accompanied by huge swings in $\theta$ near the peak of the flare. We also note that these swings were seen in all the five bands. This pattern is extremely hard to reconcile with the notion of a bremsstrahlung-dominated flare.

It is also interesting that the large swing in $\theta$ within  2 days appears superposed on a fairly steadily rising $\theta$-profile during that week when the QPOO flare stayed above the background (Fig. \ref{fig:LC}). This indicates presence of two emission components within the flare. Such a two-component structure is also discernible in the flux profile of this QPOO (Fig. \ref{fig:LC}), wherein a shoulder-like formation can be seen on 6 January, after which the flare steepens again to reach its maximum on 8 January. A similar hint is present on the declining side of the flare (Fig. \ref{fig:LC}). Thus, conceivably, a short-duration component in the flux profile, lasting for two to three days, which we associate with the simultaneously observed large to-and-fro swing in $\theta$, is superposed on a medium-duration ($\sim$ 10 days) component that may be associated with the above-mentioned steady rise in $\theta$ (Fig. \ref{fig:LC}). Another interesting point to note is that the large abrupt swing in $\theta$ over 6-9 January, particularly the sharp dip, is distinctly chromatic, becoming less pronounced on the redder side (Fig. \ref{fig:LC}, \citealp[see also][]{Holmes1984MNRAS.211..497H}).
\section{Summary and conclusions}

In this work, we scrutinise the published optical photo-polarimetric monitoring data for three large and prominent QPOOs of OJ 287, selected in an unbiased manner. These three QPOOs have been widely acclaimed in the literature as bremsstrahlung impact flares. We  subjected them to a variety of observational tests, based on the premise that the influence of any bremsstrahlung flare should be reflected in the behaviour of three measured optical parameters, namely $\alpha$, $P,$ and $\theta$. The data availability \citep[][]{Smith1987ApJS...64..459S,Sillanpaa1991Ap&SS.176..297S} enabled us to conduct all three tests for the giant QPOO of January 1983. For the remaining two large QPOOs (September 2007 and December 2015), the available data \citep[][]{Valtonen2008Natur.452..851V,Valtonen2016ApJ...819L..37V,Gupta2017MNRAS.465.4423G}  only allowed for two of the three tests to be performed (Sect. 2).  In not a single one of these seven instances of testing did we find a consonance with the thermal dilution hypothesis, which rests on the bremsstrahlung-dominated interpretation of the QPOOs during the rising phase. Perhaps the most glaring discrepancy pertains to the $\theta-$test, since for both QPOOs (January 1983 and September 2007) for which the published data have permitted this test, a statistically robust huge swing in $\theta$ was observed very close to the continuum peak. Likewise, the expected drop in $P$ in sync with the continuum peak is found to be either undetected, or inexplicably large for the thermal dilution scenario, or temporally offset from the continuum peak. Thus,  altogether, the present analysis bolsters the case for an alternative mechanism, most probably synchrotron, even for the radio-undetected flares of the QPOOs of OJ 287 (see below).

For the giant QPOO of January 1983, the present analysis has hinted at two emission components within the purported bremsstrahlung flare. These (likely synchrotron) components, lasting $\sim$ 10 days and $\sim$ 2 - 3 days, exhibit very different patterns of $\theta$ variability (Fig. \ref{fig:LC}; Sect. 2.3). \citet[][]{Holmes1984MNRAS.211..497H} sought to explain the observed large swing in $\theta$ near the flux maximum, in terms of two polarised synchrotron sources of comparable luminosities, moving in a relativistic jet, such that their spectral indices differ by $\sim$ 0.5 and their polarisation angles are mutually orthogonal, for instance, a shock forming in a stable jet flow. However, that analysis has 
somehow remained largely ignored in the extensive literature published on this blazar, \citep[see, e.g.][]{Valtonen2023aGalax..11...82V,Kushwaha2022JApA...43...79K}. 

According to a conceivable, albeit fairly speculative alternative scenario sketched below, the impact of the secondary SMBH on the accretion disk of the primary SMBH triggers jet activity in the secondary SMBH, possibly via Roche-lobe flooding \citep [e.g.][]{Pihajoki2013aMNRAS.434.3122P,Pihajoki2013bApJ...764....5P,Valtonen2023aGalax..11...82V}. However, the idea of triggering a short-lived powerful jet in the secondary SMBH during its collision with the putative thin disk would require improved theoretical and observational underpinnings, given the extremely violent circumstances expected to prevail during such collisions \citep[see e.g.][]{Komossa2022MNRAS.513.3165K}. Nonetheless, with this caveat in mind, it is tempting in the context of the January 1983 QPOO, to identify the secondary jet-triggering hypothesis with the above-mentioned first emission component found to last a few days (see Sect. 2.3). The second emission component (of longer duration) could then be an aggregate contribution coming from synchrotron pockets formed near the SMBH-impacted part of the disk which can be strongly magnetised with B $ > 10^4$ G \citep[][]{Valtonen2021Galax..10....1V}.
Theoretical studies show that the conditions for synchroron emission can arise in the strong shocks of the SMBH-disk collision \citep[see e.g.][]{Medvedev1999ApJ...526..697M,Giacalone2007AIPC..932..243G}.
Their synchrotron flux would remain subdued initially because of synchrotron self-absorption and/or free-free absorption, the latter declining as the surrounding impact-driven plasma bubble causing the absorption expands out of the disk} \citep[][]{Ivanov1998ApJ...507..131I,Valtonen2021Galax..10....1V}. 
Starting from the actual impact event, the computation of the actual time of onset of synchrotron flare (proposed to be contributed by the magnetised pockets near the impact site, as well as the nascent secondary jet) could be beset with a good deal of uncertainty of astrophysical origin. This problem is compounded for the standard DP model since for predicting the times of appearance of the impact flares (QPOOs), it invokes a bremsstrahlung-dominated post-impact astrophysics, a premise now put to serious doubt by the present analysis of three famous, well-observed QPOOs  that have hitherto been proclaimed to be outstanding cases of 'bremsstrahlung flares'. We note that these cautionary remarks become specially relevant since a sub-day precision has sometimes been estimated or claimed for a predicted time interval between the outbursts observed a decade apart \citep[see e.g.][]{Valtonen2023aGalax..11...82V,Laine2020ApJ...894L...1L}. In general, such an accuracy seems overly optimistic, considering the complex post-impact astrophysics involved, as mentioned above. Furthermore, it may be noted that even when the opacity of the expanding plasma cloud has declined enough to allow escape of the optical photons arising from the embedded synchrotron pockets and the young secondary jet, the opacity may still remain too high for their radio-frequency photons to escape. This could well explain the oft-reported radio non-detection of the initial main flare of the QPOOs (Sect. 1). Another possibility is that the very strong magnetic field near the accretion disk may often shift the bulk of the synchrotron output into the optical regime 
\citep[see][]{Valtonen2019ApJ...882...88V}. 


The work presented here covers the period until the end of 2015 when the QPOO known as the `GR flare' was observed, marking 100 years since the completion of Einstein's theory of general relativity \citep[see][]{Valtonen2016ApJ...819L..37V,Dey2018ApJ...866...11D,Valtonen2019ApJ...882...88V}. Then, in late 2015, a new observational phase of this blazar commenced under project MOMO (Sect. 1), resulting in an exceptionally dense long-term multi-band monitoring of this blazar. The largest outburst observed under this project so far had a duration of at least 0.5 yr, with brightness in different bands peaking between October 2016 and March 2017
\citep[][]{Komossa2020MNRAS.498L..35K,Komossa2023aApJ...944..177K}.  Various observational indicators favour a synchrotron origin for this high-amplitude outburst \citep[][and references therein]{Komossa2023aApJ...944..177K}. Coming $< 1$ yr after the QPOO, called the `GR flare' or the `centennial flare' mentioned above, the 2016/17 outburst was initially perceived as jet-flaring activity unrelated to the SMBH binary or a primary jet-driven `after flare' of the GR impact flare \citep[][]{Valtonen2017Galax...5...83V,Komossa2021cMNRAS.504.5575K}, as envisioned in the binary black hole model of OJ 287, wherein the impact of the secondary black hole on the disk could trigger time-delayed accretion and new jet activity in the primary black hole \citep[][]{Sundelius1997ApJ...484..180S,Valtonen2009ApJ...698..781V}. However,  it was recently suggested that in several properties, the pronounced outburst of 2016/17 matches the previous well known double-peaked impact flares of OJ 287 observed during 1970-2000; thus, it may probably be the latest addition to the long sequence of recorded optical outbursts of this blazar \citep[][]{Komossa2023aApJ...944..177K}, except that its timing is clearly inconsistent with the DP model prediction. If this surmise is correct, the optical-UV-X-ray outburst of 2016-17, which is also accompanied throughout by radio emission and by $\gamma-ray$ emission in its rising phase, would be the most recent double-peaked outburst of OJ 287, with even its  first peak being synchrotron-dominated. In that case, we note that the first peak is identifiable with a UV flare observed around February 2016 and the second peak with a much larger UV flare seen in October 2016, both having radio counterparts, as per \citet[Fig. 9][]{Komossa2023aApJ...944..177K}
\footnote{ It may be recalled that for the impact flare of OJ 287, termed `Eddington flare', observed in July 2019 in the near-infrared/optical bands, a thermal spectrum has been inferred, based on relatively sparce data \citep[][]{Laine2020ApJ...894L...1L}.} 
The next impact outburst of OJ 287, which will be verifiable in the near future, is predicted to begin over 2026-28 \citep[][]{Komossa2023aApJ...944..177K}.  Interestingly, this timing agrees with the prediction coming from a totally different approach, namely the `precessing jet' model of this blazar reported in \citet[][]{Britzen2018MNRAS.478.3199B}.

Reverting to the main theme of the present study, we have shown here that the published optical photo-polarimetric measurements of three very prominent and well-established  QPOOs of the blazar OJ 287, selected in an unbiased manner, are inconsistent with their much-acclaimed bremsstrahlung interpretation. Instead, the data distinctly favour the alternative of synchrotron mechanism for all these QPOOs. Possible reasons for the non-detection of the QPOOs at radio wavelengths are outlined. The multi-band search for future QPOO outbursts of OJ 287, expected to occur in 2026-28 (see above), should enable strong constraints to be placed on the parameters of the SMBH binary and on the specific physical mechanism behind the origin of the QPOOs. 

\section{Acknowledgements}

I thank Drs. Dusmanta Patra, Krishan Chand and Vibhore Negi for discussions and the Indian National Science Academy for a Senior Scientist position.

\bibliographystyle{aa}
\bibliography{reference}

\end{document}